\documentclass[%
 reprint,
superscriptaddress,
showpacs,
 amsmath,amssymb,
 aps,
prc
]{revtex4-1}

\usepackage{graphicx}
\usepackage{dcolumn}
\usepackage{bm}
\usepackage{amsmath}
\allowdisplaybreaks[1]
\usepackage{isotope}
\usepackage{color}

 \usepackage{changes}

\begin{document}


\title{Nuclear collective dynamics in the lattice Hamiltonian Vlasov method}

\author{Rui Wang}
\affiliation{School of Physics and Astronomy and Shanghai Key Laboratory for
Particle Physics and Cosmology, Shanghai Jiao Tong University, Shanghai $200240$, China}
\affiliation{Shanghai Institute of Applied Physics, Chinese Academy of Sciences, Shanghai $201800$, China.}
\author{Lie-Wen Chen}
\thanks{Corresponding author: lwchen$@$sjtu.edu.cn}
\affiliation{School of Physics and Astronomy and Shanghai Key Laboratory for
Particle Physics and Cosmology, Shanghai Jiao Tong University, Shanghai $200240$, China}
\author{Zhen Zhang}
\affiliation{Sino-French Institute of Nuclear Engineering and Technology,
Sun Yat-Sen University, Zhuhai $519082$, China}

\date{\today}

\begin{abstract}
The lattice Hamiltonian method is developed for solving the Vlasov equation with nuclear mean-field based on the Skyrme pseudopotential up to next-to-next-to-next-to leading order.
The ground states of nuclei are obtained through varying the total energy with respect to the density distribution of nucleons.
Owing to the self-consistent treatment of initial nuclear ground state and the exact energy conservation in the lattice Hamiltonian method, the present framework of solving the Vlasov equation exhibits very stable nuclear ground state evolution.
As a first application of the new lattice Hamiltonian Vlasov method, we explore the isoscalar giant monopole and isovector giant dipole modes of finite nuclei.
The obtained results are shown to be comparable to that from random-phase approximation and are consistent with the experimental data, indicating the capability of the present method in dealing with the long-time near-equilibrium nuclear dynamics.
\end{abstract}


\maketitle


\section{Introduction}

One-body transport models like the Boltzmann-Uehling-Uhlenbeck~(BUU) equation (see, e.g., Ref.~\cite{BerPR160}) provide a powerful tool to describe heavy-ion collisions.
Based on these models, valuable information on the details of nuclear reaction dynamics and nuclear matter properties has been obtained by analyzing data in heavy-ion collisions~\cite{BLIJMPE7,Dansc298,BarPR410,BLPR464}.
Among the obtained information, those concerning the equation of state~(EOS) of asymmetric nuclear matter at both subsaturation and suprasaturation densities and the in-medium nuclear effective interaction are of fundamental importance, since they are crucial in investigating the properties of various nuclear systems or astrophysical objects~\cite{BLIJMPE7,Dansc298,Latsc304,BarPR410,StePR411,LatPR442,BLPR464,TraIJMPE21,HorJPG41,BLEPJA50,HebARNPS65,BalPPNP91,OerRMP89,BLNPN27,BLPPNP99}.
Heavy-ion collisions not only serve as an alternative method in extracting information on subsaturation EOS through collective flows and particle production~(at intermediate beam energy)~\cite{BLPRL85,BarPR410,BLPR464,LCPRL94,TsaPRL102},
but also turn out to be the unique tool in terrestrial labs for the exploration of the suprasaturation density behaviors of the nuclear matter EOS~(at intermediate to high beam energy)~\cite{BLPRL88,XZGPRL102,FZQPLB683,RusPLB697,WXPLB718,CozPRC88,RusPRC94,CozEPJA54}.
In this sense, it is important to improve and upgrade the one-body transport models so that they can help to provide more reliable information on the EOS of asymmetric nuclear matter and the in-medium nuclear effective interaction.
To this end, the community of heavy-ion collisions at intermediate energies have actually made great efforts in recent years to improve the transport models for heavy-ion collisions by the program of transport model code comparisons~\cite{XJPRC93,ZYXPRC97}.

One of the basic input in one-body transport models is the single nucleon potential~(nuclear mean-field potential) under non-equilibrium conditions, which is generally dependent on nucleon momentum.
The momentum dependence of single nucleon potential, which has various origins like the exchange term of finite-range nuclear interaction and nuclear short-range correlations, is evident from the observed momentum or energy dependence of nucleon optical model potential~\cite{HamPRC41,CooPRC47}.
This has driven the construction and development of a lot of momentum-dependent mean-field potentials, and they have been extensively employed to study both nuclear matter and
heavy-ion collisions~\cite{GalPRC35,PraPRC37,WelPRC38,GrePRC59,DanNPA673,Bom2001,PerPRC65,DasPRC67,LCPRL94,CXPRC81,JXPRC82,LCEPJA50,JXPRC91}.
Most of momentum-dependent mean-field potentials so far applied in one-body transport models are parameterized phenomenologically and they cannot be directly used in nuclear structure calculations.
It is thus interesting and constructive to directly employ the momentum-dependent nuclear effective interactions, which are widely used in mean-field calculations of finite nuclei, to one-body transport models.
By doing this, experimental observables from finite nuclei and heavy-ion collisions can provide crosschecks for the single nucleon potentials, and thus enhance the understanding on in-medium nuclear effective interactions and the associated nuclear matter EOS.

The Skyrme interaction~\cite{SkyPM1,SkyNP9} has been used very successfully in describing the ground and lowly excited state properties of finite nuclei in mean-field calculations~\cite{BenRMP75,StoPPNP58}, as well as heavy-ion collisions at low energies in time-dependent Hartree-Fock (TDHF) calculations~\cite{MarCPC185,NakRMP88}.
However, its incorrect momentum dependence at high kinetic energies~(above about $300~\rm MeV/nucleon$), which fails to reproduce the empirical results on the nucleon optical potential obtained by Hama \textsl{et al.}~\cite{HamPRC41,CooPRC47}, hinders the application of the conventional Skyrme interaction~\cite{VauPRC5,VauPRC7,ChaNPA627,ChaNPA635} in transport model calculations for heavy-ion collisions at intermediate and high energies.
Recent development of quasi-local energy density functional and Skyrme pseudopotential~(the terminology represents effective interactions with quasi-local operators depending on spatial derivatives) makes it possible to incorporate the Skyrme interaction into the transport model calculations for intermediate to high energy heavy-ion collisions.
This is achieved by introducing additional higher order derivative terms (or higher order momentum dependence) in the conventional Skyrme effective interaction.
In Ref.~\cite{WRPRC98}, based on the next-to-next-to-next-to leading order~(N$3$LO) Skyrme pseudopotential, the Hamiltonian density and single nucleon potential under general non-equilibrium condition have been given, and three N$3$LO Skyrme pseudopotentials with single particle potential consistent with empirical optical potential up to $1~\rm GeV$ kinetic energy have been constructed.
This provides the possibility to study finite nuclei and heavy-ion collisions at incident energy up to about $1~\rm GeV/nucleon$ (where nuclear matter with about three times of saturation density can be formed~\cite{BLNPA708}) on the same footing by using the same nuclear effective interaction.

Another important aspect of developing the transport model for heavy-ion collisions is to improve the numerical stability as well as to guarantee the energy and momentum conservation during the dynamic process. The lattice Hamiltonian method~\cite{LenPRC39} provides a good recipe for this motivation. The lattice Hamiltonian method conserves the total energy exactly and the total momentum to a high degree of accuracy, and has been successfully employed into the study of heavy-ion collisions~\cite{GalPRC44,HXPRL65,HXPLB261,HXPRL67,HXPRC46a,HXPRC46b,HXPLB299,GrePRC59,PerPRC65,DasPLB726,MalPRC89,MalPRC91}.
As the first step of building a transport model with the lattice Hamiltonian method by incorporating the N$3$LO Skyrme pseudopotential, in the present work we develop a one-body transport model without considering nucleon-nucleon collisions based on the lattice Hamiltonian Vlasov~(LHV) method~\cite{LenPRC39} with nuclear mean-field potential based on the N$3$LO Skyrme pseudopotential.
In order to ensure the reliability of the LHV method, the initial phase space distribution is obtained self-consistently through varying the total energy.
The ground state properties of the present LHV transport model are examined.
Such a LHV transport model without nucleon-nucleon collision term can be applied to the study of dynamical evolution of quantum systems where the individual collisions are either inhibited by the Pauli exclusive principle or negligible due to the diluteness of the system.
In that case, collective motions of finite nuclei are ideal sites to test the validity of the present LHV method.
To this end, isoscalar giant monopole and isovector giant dipole modes of \isotope[208]{Pb} are studied based on the present LHV method.
The obtained results are compared with that from random-phase approximation~(RPA) and experimental data.

The present article is organized as follows.
In Sec.~\ref{S:SP}, we introduce the mean-field potential used in the LHV method, namely, N$3$LO Skyrme pseudopotential, and its Hamiltonian density under general nonequilibrium conditions.
In Sec.~\ref{S:LHV}, the details of the LHV method are given, and the initialization of nuclear ground state and the treatment of collective motions used in the present work are also introduced.
The ground state properties, isoscalar giant monopole, and isovector giant dipole modes of \isotope[208]{Pb} based on the LHV method with the Skyrme pseudopotentials are presented in Sec.~\ref{S:R&D}.
Finally, we summarize the present work and make a brief outlook in Sec.~\ref{S:S&O}.

\section{\label{S:SP}Mean-field potential}

\subsection{N3LO Skyrme pseudopotential}

The quasilocal nuclear energy density functional~(EDF) based on the density-matrix expansion provides an efficient way to investigate the universal EDF of a nuclear system.
In previous literature~\cite{CarPRC78,RaiPRC83}, the Skyrme interaction has been recognized as the corresponding pseudopotential of quasilocal nuclear EDF through Hartree-Fock~(HF) approximation, and a mapping has been established from the N$3$LO local EDF~\cite{CarPRC78} to N$3$LO Skyrme pseudopotential.
Such a mapping is worthwhile because it provides an order-by-order way to examine the validity of each term in the quasilocal effective interaction, since the precise structure of nuclear EDF can be derived from low-energy quantum chromodynamics with chiral perturbation theory~\cite{PugNPA723,KaiNPA724,FinNPA770}.
The N$3$LO Skyrme pseudopotential~\cite{CarPRC78,RaiPRC83} is a generalization of the standard Skyrme interaction~\cite{VauPRC5,VauPRC7,ChaNPA627,ChaNPA635} by adding terms that depend on derivative operators~(momentum operator) up to sixth order, which corresponds to the expansion of the momentum space matrix elements of a generic interaction in powers of the relative momenta up to the sixth order.
In this sense, the standard Skyrme interaction can be regarded as a N$1$LO Skyrme pseudopotential.
Such kind of generalization of the Skyrme interaction has been employed to describe EOS of nuclear matter~\cite{DavJPG40,DavJPG41,DavPS90,DavPRC91,DavAA585}, as well as the properties of finite nuclei~\cite{CarCPC181,BecPRC96}.

The full Skyrme pseudopotential generally contains central, spin-orbit, and tensor components.
Since in the present LHV method, and in most of the one-body transport models, only spin-averaged quantities are taken into consideration, we thus ignore the spin-orbit and tensor components, which are irrelevant to spin-averaged quantities.
The corresponding extended Skyrme interaction used in the present LHV method is written as
\begin{equation}\label{E:VSk}
  v_{Sk} = V^{C}_{\text{N3LO}} + V^{DD}_{\text{N1LO}},
\end{equation}
with an overall factor $\hat{\delta}(\vec{r}_1-\vec{r}_2)$ omitted for the sake of clarity.
The central term is expressed as
\begin{widetext}
\begin{align}\label{E:VSkC}
V^{C}_{\text{N3LO}} = &~t_0(1+x_0\hat{P}_{\sigma}) + t^{[2]}_1(1+x^{[2]}_1\hat{P}_{\sigma})\frac{1}{2}(\hat{\vec{k}}'^2 + \hat{\vec{k}}^2) + t^{[2]}_2(1 + x^{[2]}_2\hat{P}_{\sigma})\hat{\vec{k}}'\cdot\hat{\vec{k}} + t^{[4]}_1(1+x^{[4]}_1\hat{P}_{\sigma})\Big[\frac{1}{4}(\hat{\vec{k}}'^2+\hat{\vec{k}}^2)^2+(\hat{\vec{k}}'\cdot\hat{\vec{k}})^2\Big]\notag\\
&+ t^{[4]}_2(1+x^{[4]}_2\hat{P}_{\sigma})(\hat{\vec{k}}'\cdot\hat{\vec{k}})(\hat{\vec{k}}'^2+\hat{\vec{k}}^2) + t^{[6]}_1(1 + x^{[6]}_1\hat{P}_{\sigma})(\hat{\vec{k}}'^2 + \hat{\vec{k}}^2)\Big[\frac{1}{2}(\hat{\vec{k}}'^2 + \hat{\vec{k}}^2)^2 + 6(\hat{\vec{k}}'\cdot\hat{\vec{k}})^2\Big]\notag\\
&+ t^{[6]}_2(1+x^{[6]}_2\hat{P}_{\sigma})(\hat{\vec{k}}'\cdot\hat{\vec{k}})[3(\hat{\vec{k}}'^2+\hat{\vec{k}}^2)^2+4(\hat{\vec{k}}'\cdot \hat{\vec{k}})^2],
\end{align}
\end{widetext}
where $\hat{\vec{k}}'$ and $\hat{\vec{k}}$ are derivative operators acting on left and right, and they take the conventional form as $-(\hat{\vec{\nabla}}_1-\hat{\vec{\nabla}}_2)/2i$ and
$(\hat{\vec{\nabla}}_1-\hat{\vec{\nabla}}_2)/2i$, respectively.
$\hat{P}_{\sigma}$ represents the spin exchange operator defined as $\hat{P}_{\sigma} = \frac{1}{2}(1 + \hat{\sigma}_1\hat{\sigma}_2)$, where $\hat{\sigma}_1$ and $\hat{\sigma}_2$ are Pauli matrices acting on the first and second state, respectively.
The density-dependent term $V^{DD}_{\text{N1LO}}$, which is introduced to mimic phenomenologically the effects of many-body interactions, is taken to be the same form as in the standard Skyrme interaction, i.e.,
\begin{equation}\label{E:VSkDD}
  V^{DD}_{\text{N1LO}} = \frac{1}{6}t_3(1+x_3\hat{P}_{\sigma})\rho^{\alpha}\Big(\frac{\vec{r}_1 +\vec{r}_2}{2}\Big).
\end{equation}
In the above expressions, $t^{[n]}_i$, $x^{[n]}_i$ ($n=2, 4, 6$ and $i = 1, 2$), $t_0$, $t_3$, $x_0$, $x_3$ and $\alpha$ are Skyrme parameters. In particular, the parameters $t_0$ and $t^{[n]}_i$ are measures of the mean central potential, with an spin-exchange character specified by $x_0$ and $x^{[n]}_i$.
The parameters $t_3$ and $x_3$ are their analogs of the density-dependent potential, with $\alpha$ characterizing its density dependence.
The parameters $t^{[n]}_i$ and $x^{[n]}_i$ are related to the derivative terms which determine the momentum-dependent parts of the Hamiltonian density and the single-nucleon potentials.
With the introducing of the additional derivative terms in Eq.~(\ref{E:VSkC}), the N$3$LO Skyrme pseudopotential is able to describe the empirical single nucleon potential up to a kinetic energy of $1~\rm GeV$\cite{WRPRC98}.

\subsection{\label{S:H}Hamiltonian density with N$3$LO Skyrme pseudopotential}

During the heavy-ion collision process, the dinuclear system is generally far from
equilibrium.
In one-body transport models, such nonequilibrium conditions are described by the phase space distribution function (Wigner function) $f_{\tau}(\vec{r},\vec{p})$, with $\tau$ $=$ $1$~(or n) for neutrons and $-1$~(or p) for protons.
In the LHV method, we thus need to express the Hamiltonian density ${\cal H}(\vec{r})$ of the collision system in terms of $f_{\tau}(\vec{r},\vec{p})$.
The Hamiltonian density in HF approximation is obtained by calculating the expectation value of the total energy of the collision system.
For N$3$LO Skyrme pseudopotential expressed in Eq.~(\ref{E:VSk}), it takes the following form~(detailed derivation can be found in Ref.~\cite{WRPRC98}):
\begin{equation}\label{E:H}
\begin{split}
    {\cal H}(\vec{r}) & = {\cal H}^{\rm kin}(\vec{r}) + {\cal H}^{\rm loc}(\vec{r})\\
    & + {\cal H}^{\rm MD}(\vec{r}) + {\cal H}^{\rm grad}(\vec{r}) + {\cal H}^{\rm DD}(\vec{r}),
\end{split}
\end{equation}
with ${\cal H}^{\rm kin}(\vec{r})$, ${\cal H}^{\rm loc}(\vec{r})$, ${\cal H}^{\rm MD}(\vec{r})$, ${\cal H}^{\rm grad}(\vec{r})$, and ${\cal H}^{\rm DD}(\vec{r})$ being the kinetic, local, momentum-dependent, gradient, and density-dependent terms, respectively.
The kinetic term
\begin{equation}\label{E:Hkin}
    {\cal H}^{\rm kin}(\vec{r}) = \sum_{\tau = n,p}\int d^3p\frac{p^2}{2m_{\tau}}f_{\tau}(\vec{r},\vec{p})
\end{equation}
and the local term
\begin{equation}\label{E:Hloc}
    {\cal H}^{\rm loc}(\vec{r}) = \frac{t_0}{4}\bigg[(2 + x_0)\rho(\vec{r})^2 - (2x_0 + 1)\sum_{\tau = n,p}\rho_{\tau}(\vec{r})^2\bigg]
\end{equation}
are the same as those from the conventional Skyrme interaction.
The $\rho_{\tau}(\vec{r})$ in the local term is the nucleon density, which is related to
$f_{\tau}(\vec{r},\vec{p})$ through $\rho_{\tau}(\vec{r})$ $=$ $\int f_{\tau}(\vec{r},\vec{p})d\vec{p}$, while $\rho(\vec{r})$ represents the total nucleon density with $\rho(\vec{r})$ $=$ $\rho_n(\vec{r})$ $+$ $\rho_p(\vec{r})$.

The momentum-dependent term and gradient term contain the contributions from additional
derivative terms in Eq.(\ref{E:VSkC}). The momentum-dependent term can be expressed as
\begin{equation}\label{E:HMD}
\begin{split}
    {\cal H}^{\rm MD}(\vec{r}) & = \int d^3pd^3p'{\cal K}_s(\vec{p},\vec{p}')f(\vec{r},\vec{p})f(\vec{r},\vec{p}')\\
    & + \sum_{\tau = n, p}\int d^3pd^3p'{\cal K}_v(\vec{p},\vec{p}')f_{\tau}(\vec{r},\vec{p})f_{\tau}(\vec{r},\vec{p}'),
\end{split}
\end{equation}
with $f(\vec{r},\vec{p})$ $=$ $f_n(\vec{r},\vec{p})$ $+$ $f_p(\vec{r},\vec{p})$.
The ${\cal K}_{\rm s}(\vec{p},\vec{p}')$ and ${\cal K}_{\rm v}(\vec{p},\vec{p}')$ in Eq.~(\ref{E:HMD}) represent the isoscalar and isovector momentum-dependent kernel of mean-field potential, respectively.
For N$3$LO Skyrme pseudopotential used in the present work, ${\cal K}_{\rm s}(\vec{p},\vec{p}')$ and ${\cal K}_{\rm v}(\vec{p},\vec{p}')$ take the following forms,
\begin{align}
    {\cal K}_{\rm s}(\vec{p},\vec{p}') & = \frac{C^{[2]}}{16\hbar^2}(\vec{p} - \vec{p}')^2 + \frac{C^{[4]}}{32\hbar^2}(\vec{p} - \vec{p}')^4\notag\\
    & + \frac{C^{[6]}}{16\hbar^2}(\vec{p} - \vec{p}')^6,\label{E:mdks}\\
    {\cal K}_{\rm v}(\vec{p},\vec{p}') & = \frac{D^{[2]}}{16\hbar^2}(\vec{p} - \vec{p}')^2 + \frac{D^{[4]}}{32\hbar^2}(\vec{p} - \vec{p}')^4\notag\\
    & + \frac{D^{[6]}}{16\hbar^2}(\vec{p} - \vec{p}')^6.\label{E:mdkv}
\end{align}
In the present work, for the sake of simplicity of numerical derivatives, the gradient term is truncated at the second order, and only the isospin symmetric part is taken into account.
In other words, we only keep the second order derivative of the total baryon density $\rho(\vec{r})$, i.e.,
\begin{equation}\label{E:Hgrad}
    {\cal H}^{\rm grad}(\vec{r}) = \frac{1}{16}E^{[2]}\Big\{2\rho(\vec{r})\nabla^2\rho(\vec{r}) - 2\big[\nabla\rho(\vec{r})\big]^2\Big\}.
\end{equation}
The complete gradient term in the Hamiltonian density for the N$3$LO Skyrme potential can be found in Ref.~\cite{WRPRC98}.

In the above expressions, for convenience, we have recombined the Skyrme parameters
related to the derivative terms in Eq.~(\ref{E:VSkC}), namely, $t_1^{[n]}$, $t_2^{[n]}$,
$x_1^{[n]}$, and $x_2^{[n]}$, into the parameters $C^{[n]}$ and $D^{[n]}$,
\begin{align}
C^{[n]} & = t_1^{[n]}(2+x_1^{[n]})+t_2^{[n]}(2+x_2^{[n]}),\label{E:Cn}\\
D^{[n]} & = -t_1^{[n]}(2x_1^{[n]}+1)+t_2^{[n]}(2x_2^{[n]}+1),\label{E:Dn}
\end{align}
which are related to the momentum-dependent terms, and $E^{[2]}$, which is related to the gradient terms with
\begin{equation}
    E^{[2]} = -\frac{1}{4}\big[t_1^{[2]}(2+x_1^{[2]}) - t_2^{[2]}(2+x_2^{[2]})\big].\label{E:E2}
\end{equation}
The density-dependent term is expressed as
\begin{equation}\label{E:HDD}
\begin{split}
    {\cal H}^{\rm DD}(\vec{r}) = \frac{t_3}{24}\bigg[(2 + x_3)\rho^2 - (2x_3 + 1)\sum_{\tau=n,p}\rho_{\tau}^2\bigg]\rho^{\alpha}.
\end{split}
\end{equation}
Based on the above expressions, one can see that the Hamiltonian density ${\cal H}(\vec{r})$ is explicitly dependent on $f_{\tau}(\vec{r},\vec{p})$, as well as the $\rho_{\tau}(\vec{r})$ and their derivatives.

\begin{table}[!htb]
\centering
\caption{Parameters related to the nuclear matter properties of three N$3$LO Skyrme pseudopotentials, SP$6$s, SP$6$m, and SP$6$h, and one conventional Skyrme interaction MSL$1$, where the recombination of Skyrme parameters defined in Eqs.~(\ref{E:Cn}) and (\ref{E:Dn}) are used.}
\begin{tabular}{ccccc}
\hline\hline
 ~ & SP$6$s & SP$6$m & SP$6$h & MSL$1$\\
\hline
 $t_0$~($\rm{MeV\cdot fm}^{3}$) & -1814.64 & -1956.75 & -1675.52 & 1963.23\\
 $x_0$ & 0.5400 & 0.2306 & -0.0902 & 0.3208\\
 $t_3$~($\rm{MeV\cdot fm}^{3 + 3\alpha}$) & 10796.2 & 11402.9 & 9873.1 & 12174.9\\
 $x_3$ & 0.8257 & 0.1996 & -0.4990 & 0.3219\\
 $\alpha$ & 0.2923 & 0.2523 & 0.3168 & 0.2694\\
 $C^{[2]}$~($\rm{MeV\cdot fm}^{5}$) & 597.877 & 637.195 & 677.884 & 435.519\\
 $D^{[2]}$~($\rm{MeV\cdot fm}^{5}$) & -446.695 & -524.373 & -601.990 & -367.583\\
 $C^{[4]}$~($\rm{MeV\cdot fm}^{7}$) & -26.2027 & -28.5209 & -31.2026 & 0.0\\
 $D^{[4]}$~($\rm{MeV\cdot fm}^{7}$) & 23.2525 & 27.6873 & 32.4607 & 0.0\\
 $C^{[6]}$~($\rm{MeV\cdot fm}^{9}$) & 0.0903 & 0.1000 & 0.1121 & 0.0\\
 $D^{[6]}$~($\rm{MeV\cdot fm}^{9}$) & -0.0896 & -0.1080 & -0.1292 & 0.0\\
\hline\hline
\end{tabular}
\label{T:SPs}
\end{table}

The calculations in the present work are mainly based on three N$3$LO Skyrme pseudopotentials, SP$6$s, SP$6$m and SP$6$h~\cite{WRPRC98}, which can describe the empirical single nucleon potential up to $1~\rm GeV$ in kinetic energy.
The main difference of these three interactions is their suprasaturation behavior of the isospin dependent part of EOS, namely, the symmetry energy.
Since the nucleon momenta in both the ground state properties and low energy collective excitation are not much larger than the Fermi momentum of saturated nuclear matter, conventional Skyrme interactions are still able to give the empirical single nucleon potential.
In order to compare with RPA calculations, one conventional Skyrme interaction MSL$1$~\cite{ZZPLB726} is also adopted in the present work.
We list the Skyrme parameters related to nuclear matter properties for the above four Skyrme interactions in Table~\ref{T:SPs}, while more discussions about the gradient parameter $E^{[2]}$, which is irrelevant to nuclear matter properties, will be given in Sec.~\ref{S:INI}.
For further reference, the characteristic parameters of nuclear matter for these interactions are shown in Table~\ref{T:CPs}.
The definition of these quantities can be found, e.g., in Ref.~\cite{LCPRC80}.

\begin{table}[!htb]
\centering
\caption{Macroscopic characteristic quantities for SP$6$s, SP$6$m, SP$6$h, and MSL$1$, where $\rho_{\rm sc}$ $=$ $0.11/0.16\rho_0$ and $\rho_{\rm h}$ $=$ $0.5~{\rm fm}^{-3}$.}
\begin{tabular}{ccccc}
\hline\hline
 ~ & SP$6$s & SP$6$m & SP$6$h & MSL$1$\\
 \hline
 $\rho_{0}$~($\rm{fm}^{-3}$) & $0.1614$ & $0.1630$ & $0.1647$ & $0.1586$\\
 $E_{0}$~($\rm{MeV}$) & $-16.04$ & $-15.94$ & $-15.61$ & $-16.00$\\
 $K_{0}$~($\rm{MeV}$) & $240.9$ & $233.4$ & $240.8$ & $235.1$\\
 $J_0$~($\rm{MeV}$) & $-377.0$ & $-384.2$ & $-358.2$ & $-372.7$\\
 $E_{\rm{sym}}(\rho_{\rm{sc}})$~($\rm{MeV}$) & $25.43$ & $25.83$ & $25.98$ & $26.67$\\
 $L(\rho_{\rm sc})$~($\rm MeV$) & $32.47$ & $46.75$ & $62.19$ & $46.19$\\
 $E_{\rm sym}(\rho_0)$~($\rm MeV$) & $28.84$ & $31.93$ & $34.97$ & $32.33$\\
 $L(\rho_0)$~($\rm MeV$) & $18.20$ & $49.10$ & $82.17$ & $45.25$\\
 $K_{\rm sym}$~($\rm MeV$) & $-242.7$ & $-158.0$ & $-70.5$ & $-183.3$\\
 $E_{\rm sym}(2\rho_0)$~($\rm MeV$) & $ 24.06 $ & $ 41.31 $ & $ 61.62 $ & $39.00$\\
 $E_{\rm sym}(\rho_{\rm h})$~($\rm MeV$) & $0.03$ & $41.32$ & $79.82$ & $31.01$\\
 $m_{s,0}^{\ast}/m$ & $0.759$ & $0.758$ & $0.755$ & $0.806$\\
 $m_{v,0}^{\ast}/m$ & $0.678$ & $0.663$ & $0.648$ & $0.706$\\
\hline\hline
\end{tabular}\label{T:CPs}
\end{table}

\section{\label{S:LHV}Lattice Hamiltonian Vlasov method}
\subsection{Lattice Hamiltonian method for Vlasov equation}
Quantum theory with phase-space distributions, with proper generalization or approximation, is profoundly suitable to formulate and solve many-particle dynamics~\cite{CarRMP55}.
It is demonstrated that in the limit involving $\hbar$ $\rightarrow$ $0$, quantum theory with one-body phase-space distributions is reduced to the Vlasov equation~\cite{BerPR160},
\begin{equation}\label{E:VE}
    \frac{\partial f}{\partial t} + \frac{\vec{p}}{E}\nabla_{\vec{r}}f + \nabla_{\vec{p}}U(\vec{r},\vec{p})\cdot\nabla_{\vec{r}}f - \nabla_{\vec{r}}U(\vec{r},\vec{p})\cdot\nabla_{\vec{p}}f = 0,
\end{equation}
where $f$ is the one-body phase-space distribution, or Wigner function defined as the Wigner transform of one-body density matrix $\rho(\vec{r}+\vec{s}/2,\vec{r}-\vec{s}/2)$, i.e.,
\begin{equation}
    f(\vec{r},\vec{p}) = \frac{1}{(2\pi\hbar)^3}\int {\rm exp}\Big(-i\frac{\vec{p}}{\hbar}\cdot\vec{s}\Big)\rho(\vec{r}+\vec{s}/2,\vec{r}-\vec{s}/2)d\vec{s}.
\end{equation}
In nuclear physics, Eq.~(\ref{E:VE}) with an additional nucleon-nucleon collision term on the right hand side, which takes into account Fermi statistics, i.e.,
\begin{equation}
\begin{split}
    I_{\rm c} & = -\int\frac{d\vec{p}_2}{(2\pi\hbar)^3}\frac{d\vec{p}_3}{(2\pi\hbar)^3}\frac{d\vec{p}_4}{(2\pi\hbar)^3}|{\cal M}_{12\rightarrow34}|^2\\
    &\times(2\pi)^4\delta^4(p_1 + p_2 - p_3 - p_4)\\
    &\times[f_1f_2(1-f_3)(1-f_4) - f_3f_4(1 - f_1)(1 - f_2)],
\end{split}\label{E:Ic}
\end{equation}
is commonly referred to as the BUU equation~\cite{BerPR160}.

The Vlasov equation or BUU equation is normally solved by interpreting $f_{\tau}(\vec{r},\vec{p},t)$ as the semi-classical phase space distribution function.
If we treat each volume element as nuclear matter, the (quasi)nucleons inside the volume are in momentum eigenstates obeying the Pauli principle.
Under such a condition, the obtained $f_{\tau}(\vec{r},\vec{p},t)$ through Wigner transform of the density matrix turns out to be the occupation probability of the momentum eigenstates, and thus can be regarded as the classical phase space distribution function.

The test particle method~\cite{CWPRC25}, where the semiclassical $f_{\tau}(\vec{r},\vec{p},t)$ is mimicked by a large number of test particles, i.e.,
\begin{equation}\label{E:fTP}
    f_{\tau}(\vec{r},\vec{p},t) \propto\sum_i\delta\big[\vec{r}_i(t) - \vec{r}\big]\delta\big[\vec{p}_i(t) - \vec{p}\big],
\end{equation}
has been introduced to solve the Vlasov equation numerically.
In the conventional test particle method, the evolutions of coordinate $\vec{r}_i(t)$ and momentum $\vec{p}_i(t)$ of the $i$th test nucleon are governed by the mean-field potential, or the single particle potential, which can be obtained either by varying the Hamiltonian density or by direct parametrization.
In order to obtain a smooth mean-field potential, the density of a certain cell is averaged by neighboring cells.
This kind of simple smoothing technique violates the equation of motion a little bit, and fails to conserve either the total energy or the total momentum~\cite{BerPR160}.

The lattice Hamiltonian Vlasov method developed by Lenk and Pandharipande~\cite{LenPRC39} overcomes this disadvantage of conventional test particle method.
It conserves the total energy exactly and the total momentum to a high degree of accuracy.
In the LHV method, instead of mean-field potential, the equation of motion of test nucleons is governed directly by the total Hamiltonian of the system, which is approximated by the lattice Hamiltonian, i.e.,
\begin{equation}
    H = \int {\cal H}(\vec{r})d\vec{r} \approx l_xl_yl_z\sum_{\alpha}{\cal H}(\vec{r}_{\alpha})\equiv H_L,
\end{equation}
where $\vec{r}_{\alpha}$ represents the coordinate of a certain lattice site $\alpha$ and $l_x$, $l_y$, and $l_z$ are lattice spacing.
The above lattice Hamiltonian can be expressed in terms of the positions and momenta of test nucleons, if we write the semi-classical LHV phase space distribution at lattice site $\alpha$ as
\begin{equation}\label{E:f}
    \tilde{f}_{\tau}(\vec{r}_{\alpha},\vec{p},t) = \frac{1}{2}\frac{(2\pi\hbar)^3}{N_{\rm E}}\sum_i^{\alpha,\tau}S\big[\vec{r}_i(t) - \vec{r}_{\alpha}\big]\delta\big[\vec{p}_i(t) - \vec{p}\big],
\end{equation}
where the factor $\frac{1}{2}$ is due to spin degeneracy.
$N_{\rm E}$ is the number of ensembles (or number of test particles in some literature) introduced in the calculation, and the sum runs over all test nucleons with isospin $\tau$ that contribute to the lattice site $\alpha$.
This equivalently gives each test particle a form factor $S$ compared with Eq.~(\ref{E:fTP}).
By doing this, the movement of a test particle leads to a continuous variation of the local nucleon density of the nearby lattice sites, which is useful to smooth the nucleon distribution functions in phase space.
It should be noted that the form factor $S$ actually modifies the relation between test particles and the Wigner function $f$.
At this point, we would like to point out that in principle, a similar form factor in momentum space can also be introduced in Eq.~(\ref{E:f}), which is expected to improve the calculations with momentum-dependent mean-field potentials.
In the present work, we only adopt the form factor in coordinate space, and in the future it would be definitely interesting to perform a systematic investigation on the effects of a form factor in momentum space in heavy-ion transport model calculations.
The local nucleon density at lattice sites, or LHV density, is then given by integrating $\tilde{f}_{\tau}(\vec{r}_{\alpha},\vec{p},t)$ with respect to momentum, i.e.,
\begin{equation}\label{E:rhoL}
    \tilde{\rho}_{\tau}(\vec{r}_{\alpha},t) = 2\int\tilde{f}_{\tau}\frac{d\vec{p}}{(2\pi\hbar)^3} = \frac{1}{N_{\rm E}}\sum_i^{\alpha,\tau}S\big[\vec{r}_i(t) - \vec{r}_{\alpha}\big].
\end{equation}
Note that here we distinguish the realistic phase space distribution function $f(\vec{r},\vec{p})$ and local density $\rho(\vec{r})$ from the LHV phase space distribution function and density expressed in Eqs.~(\ref{E:f}) and (\ref{E:rhoL}), respectively, and will explore their distinctions in Sec.~\ref{S:INI}.
Substituting Eq.~(\ref{E:f}) into Eq.~(\ref{E:H}), 
the lattice Hamiltonian $H_L$ is then expressed in terms of the coordinates and momenta of test nucleons, and subsequently they can be treated as the canonical variables of the lattice Hamiltonian.
The equation of motion for the $i$th test nucleon is then governed by the Hamilton equation of total lattice Hamiltonian of all ensembles $N_{\rm E}H_L$, i.e.,
\begin{widetext}
\begin{align}
\frac{d\vec{r}_i}{dt} & = N_{\rm E}\frac{\partial H_L\big[\vec{r}_1(t),\cdots,\vec{r}_{A\times N_{\rm E}}(t);\vec{p}_1(t),\cdots,\vec{p}_{A\times N_{\rm E}}(t)\big]}{\partial\vec{p}_i} = \frac{\vec{p}_i(t)}{m} + N_{\rm E}l_xl_yl_z\sum_{\alpha\in V_i}\frac{\partial{\tilde{\cal H}}^{\rm MD}_{\alpha}}{\partial\vec{p}_i}\label{E:ri},\\
\frac{d\vec{p}_i}{dt} & = - N_{\rm E}\frac{\partial H_L\big[\vec{r}_1(t),\cdots,\vec{r}_{A\times N_{\rm E}}(t);\vec{p}_1(t),\cdots,\vec{p}_{A\times N_{\rm E}}(t)\big]}{\partial\vec{r}_i}\notag\\
& = - N_{\rm E}l_xl_yl_z\sum_{\alpha\in V_i}\bigg[\sum_{\tau}^{n,p}\Big(\frac{\partial{\tilde{\cal H}}^{\rm loc}_{\alpha}}{\partial\tilde{\rho}_{\tau,\alpha}} + \frac{\partial{\tilde{\cal H}}^{\rm Cou}_{\alpha}}{\partial\tilde{\rho}_{\tau,\alpha}} + \frac{\partial{\tilde{\cal H}}^{\rm DD}_{\alpha}}{\partial\tilde{\rho}_{\tau,\alpha}} + \frac{\partial{\tilde{\cal H}}^{\rm grad}_{\alpha}}{\partial\tilde{\rho}_{\tau,\alpha}} + \sum_n(-1)^n\nabla^n\frac{\partial{\tilde{\cal H}}^{\rm grad}_{\alpha}}{\partial\nabla^n\tilde{\rho}_{\tau,\alpha}} \Big)\frac{\partial\tilde{\rho}_{\tau,\alpha}}{\partial\vec{r}_i} + \frac{\partial{\tilde{\cal H}}^{\rm MD}_{\alpha}}{\partial\vec{r}_i}\bigg]\label{E:pi}.
\end{align}
In the above two equations, $A$ is the nucleon number of the system, while the subscripts $\alpha$ for various quantities denote their values at lattice site $\alpha$.
The sums run over all lattice sites inside $V_i$, where the form factor of the $i$th test nucleon covers.
A tilde above the Hamiltonian density, e.g., $\tilde{\cal H}^{\rm loc}(\vec{r}_{\alpha})$, denotes that in its expression in Sec.~\ref{S:H}, the realistic phase space distribution function and local density are replaced by the LHV phase space distribution function and density.
The Coulomb interaction contributes to the Hamiltonian density through
\begin{equation}
    {\cal H}^{\rm Cou}(\vec{r}_{\alpha}) = e^2\rho_p(\vec{r}_{\alpha})\bigg\{\frac{1}{2}\int\frac{\rho_p(\vec{r}')}{|\vec{r}_{\alpha} - \vec{r}'|}d\vec{r}' - \frac{3}{4}\Big[\frac{3\rho_p(\vec{r}_{\alpha})}{\pi}\Big]^{1/3}\bigg\}\label{E:Cou}\approx\tilde{\cal H}^{\rm Cou}(\vec{r}_{\alpha}) = e^2\tilde{\rho}_p(\vec{r}_{\alpha})\bigg\{\frac{1}{2}\sum_{\alpha'\ne\alpha}\frac{\tilde{\rho}_p(\vec{r}_{\alpha'})l_xl_yl_z}{|\vec{r}_{\alpha} - \vec{r}_{\alpha'}|} - \frac{3}{4}\Big[\frac{3\tilde{\rho}_p(\vec{r}_{\alpha})}{\pi}\Big]^{1/3}\bigg\},
\end{equation}
among which the minus term represents the contribution from the exchange term of Coulomb energy.
Further testing shows that the Coulomb energy $\tilde{\cal H}^{\rm Cou}(\vec{r}_{\alpha})$ defined in the above equation has already converged at lattice spacing of $l_x$ $=$ $l_y$ $=$ $l_z$ $=$ $0.5~\rm fm$ used in the present work.
The partial derivative of $\tilde{\rho}_{\tau,\alpha}$ in Eq.~(\ref{E:pi}) can be calculated in terms of the spatial derivative of $S$, i.e.,
\begin{equation}
    \frac{\partial\tilde{\rho}_{\tau,\alpha}}{\partial\vec{r}_i} = \frac{\partial}{\partial\vec{r}_i}\sum_{\vec{r}_j\in V_{\alpha}}^{\tau_j=\tau}S(\vec{r}_j-\vec{r}_{\alpha})
     = \begin{cases}
         & \frac{\partial S(\vec{r}_i-\vec{r}_{\alpha})}{\partial\vec{r}_i},\quad \tau_i = \tau,\\
         & 0,\quad \tau_i \ne \tau.
        \end{cases}
\end{equation}

The momentum-dependent parts of the equation of motion of test particles in Eqs.~(\ref{E:ri}) and (\ref{E:pi}) are obtained by substituting the momentum-dependent part of the Hamiltonian density, i.e., Eq.~(\ref{E:HMD}) into Eqs.~(\ref{E:ri}) and (\ref{E:pi}) after replacing $f_{\tau}(\vec{r},\vec{p})$ in Eq.~(\ref{E:HMD}) with the semi-classical LHV phase space distribution expressed in Eq.~(\ref{E:f}).
The integrals in Eq.~(\ref{E:HMD}) turn out to be summations of test particles,
\begin{align}
    \frac{\partial\tilde{\cal H}^{\rm MD}(\vec{r}_{\alpha})}{\partial\vec{r}_i} & =  2\frac{\partial S\big[\vec{r}_i(t) - \vec{r}_{\alpha}\big]}{\partial\vec{r}_i}\bigg\{\sum_{j\in V_{\alpha}}S\big[\vec{r}_j(t) - \vec{r}_{\alpha}\big]{\cal K}_{\rm s}\big[\vec{p}_i(t),\vec{p}_j(t)\big] + \sum_{j\in V_{\alpha}}^{\tau_j = \tau_i}S\big[\vec{r}_j(t) - \vec{r}_{\alpha}\big]{\cal K}_{\rm v}\big[\vec{p}_i(t),\vec{p}_j(t)\big]\bigg\}\label{E:mdr},\\
    \frac{\partial\tilde{\cal H}^{\rm MD}(\vec{r}_{\alpha})}{\partial\vec{p}_i} & =  2S\big[\vec{r}_i(t) - \vec{r}_{\alpha}\big]\bigg\{\sum_{j\in V_{\alpha}}S\big[\vec{r}_j(t) - \vec{r}_{\alpha}\big]\frac{\partial{\cal K}_{\rm s}\big[\vec{p}_i(t),\vec{p}_j(t)\big]}{\partial\vec{p}_i} + \sum_{j\in V_{\alpha}}^{\tau_j = \tau_i}S\big[\vec{r}_j(t) - \vec{r}_{\alpha}\big]\frac{\partial{\cal K}_{\rm v}\big[\vec{p}_i(t),\vec{p}_j(t)\big]}{\partial\vec{p}_i}\bigg\}\label{E:mdp}.
\end{align}
\end{widetext}
Based on Eqs.~(\ref{E:ri}) - (\ref{E:mdp}), we can calculate the time evolution $\vec{r}_i(t)$ and $\vec{p}_i(t)$ of test nucleons, and then calculate physical observables based on Eq.~(\ref{E:f}).

In the present work, the factor $S(\vec{r}_i - \vec{r})$ is chosen to be triangle form,
\begin{align}
    S(\vec{r}_i - \vec{r}) & = \frac{1}{(nl/2)^6}g(\Delta x)g(\Delta y)g(\Delta z),\\
    g(q) & = \Big(\frac{nl}{2} - |q|\Big)\theta\Big(\frac{nl}{2} - |q|\Big),
\end{align}
where $\theta$ is the Heaviside function, and $n$ is an integer which determines the range of $S$.
Calculations based on lattices generally break Galilean invariance, and thus violate momentum conservation.
In the present work we choose $n$ $=$ $4$, which is large enough to conserve the total momentum to a high degree of accuracy~\cite{LenPRC39}.

Generally speaking, the choice of $S(\vec{r}_i - \vec{r})$ is somewhat arbitrary.
Besides the triangle form used in the present work, alternative forms used in previous literature are trapezoid~\cite{DanNPA673}, double parabolic~\cite{PerPRC65}, and Gaussian~\cite{UrbPRC85} form.
However, in order to ensure particle number conservation, $S(\vec{r}_i - \vec{r})$ should satisfy the following equation:
\begin{equation}
    \sum_{\alpha}\tilde{\rho}(\vec{r}_{\alpha})l_xl_yl_z = \frac{1}{N_{\rm E}}\sum_{\alpha}\sum_{i}S(\vec{r}_i - \vec{r}_{\alpha})l_xl_yl_z = A.
\end{equation}

It should be mentioned that in the conventional test particle method, the Hamiltonian equations of motion for the test particles are derived from the {\it single particle} Hamiltonian, which makes it difficult to exactly conserve energy of the system in the dynamic process~\cite{BerPR160,LenPRC39}. On the other hand, in the LHV method, the Hamiltonian equations of motion for the test particles, namely, Eqs.~(\ref{E:ri}) and (\ref{E:pi}), are derived from the {\it total} Hamiltonian of the system of test particles, which guarantees the exact energy conservation of the system in the dynamic process~\cite{LenPRC39}.
In addition, we would like to mention that the present LHV method is implemented based on GPU parallel computing~\cite{Rue2013}, which increases the computational efficiency and makes it possible to obtain more reliable results with the use of much more ensembles.

\subsection{\label{S:INI}Initialization of nuclear ground state}
In the present framework, the Vlasov ground state of nuclei at zero temperature is obtained by varying the Hamiltonian with respect to the nuclear radial density.
Such kind of initialization in one-body transport model is sometimes referred to as Thomas-Fermi~(TF) initialization~\cite{LenPRC39,DanNPA673,GaiPRC81,LHPRC99}.
Within the one-body transport models, at zero temperature, for a nucleus in ground state, its Wigner function satisfies
\begin{equation}\label{E:f0}
    f_{\tau}(\vec{r},\vec{p}) = \frac{2}{(2\pi\hbar)^3}\theta\big[|\vec{p}| - p^F_{\tau}(\vec{r})\big],
\end{equation}
where $p^F_{\tau}(\vec{r})$ is local Fermi momentum and fulfills
\begin{equation}\label{E:pF}
    p^F_{\tau}(\vec{r}) = \hbar\big[3\pi^2\rho_{\tau}(\vec{r})\big]^{1/3}.
\end{equation}
For simplicity in the following we assume the nucleus is spherical.
The total energy of a ground state nucleus at zero temperature can be then treated as a functional of radial density $\rho(r)$ and its spatial gradients $\nabla_r^n\rho(r)$, i.e.,
\begin{equation}
    E = \int{\cal H}\big[r,\rho_{\tau}(r),\nabla{\rho_{\tau}(r)},\nabla^2{\rho_{\tau}(r)}\cdots\big]dr.
\end{equation}
After varying the total energy with respect to $\rho_{\tau}(r)$ and its spatial gradients, and considering Eqs.~(\ref{E:H}) (\ref{E:f0}), we obtain the neutron/proton radial density in a ground state nucleus~(note that the contribution from the Coulomb interaction in Eq.~(\ref{E:Cou}) for protons should also be included in the Hamiltonian density),
\begin{equation}\label{E:GS}
    \frac{1}{2m}\big\{p_{\tau}^F\big[\rho_{\tau}(r)\big]\big\}^2 + U_{\tau}\big\{p_{\tau}^{\rm F}\big[\rho_{\tau}(r)\big],r\big\} = \mu_{\tau},
\end{equation}
where $\mu_{\tau}$ is the chemical potential of proton or neutron inside the nucleus and $U_{\tau}\big\{p_{\tau}^{\rm F}\big[\rho_{\tau}(r)\big],r\big\}$ is the single nucleon potential of the nucleon at the Fermi surface.
The single nucleon potential is defined as the variation of Hamiltonian density with respect to the phase space distribution function~(or local density for the zero temperature case) and density gradients.
For the N$3$LO Skyrme pseudopotential, the detailed expression can be found in Ref.~\cite{WRPRC98}.
Equation (\ref{E:GS}) has a very intuitive physical significance: in a classical point of view, it means within a ground state nucleus, the nucleons possessing different Fermi momentum in different radial position have the same chemical potential.
The local density $\rho(\vec{r})$ for ground state spherical nucleus is obtained by solving Eq.~(\ref{E:GS}) subjecting to the boundary condition
\begin{equation}
    \frac{\partial\rho(r)}{\partial r}\Big|_{r = 0} = \frac{\partial\rho(r)}{\partial r}\Big|_{r = r_{\rm B}} = 0,
\end{equation}
where $r_{\rm B}$ is the boundary of the nucleus satisfying $\rho(r_{\rm B})$ $=$ $0$, and needs to be determined when solving Eq.~(\ref{E:GS}).
Since the way we deal with the ground state is semiclassical, the gradient parameter $E^{[2]}$ is readjusted for each interaction in order that the solution of Eq.~(\ref{E:GS}) reproduces roughly the experiment binding energy and charge radius of \isotope[208]{Pb}.
For MSL$1$ interaction, the results in Sec.~\ref{S:R&D} are based on the readjusted gradient parameters.

In the present LHV method, the initial coordinates of test nucleons are generated according to the solution of Eq.~(\ref{E:GS}), while their initial momenta follow zero-temperature Fermi distribution with the Fermi momentum in Eq.~(\ref{E:pF}) determined by local density.

Careful readers might realize that the obtained ground state LHV density $\tilde{\rho}(\vec{r})$ is smeared due to the form factor introduced in Eq.~(\ref{E:f}), and thus slightly different from the solution of Eq.~(\ref{E:GS}).
Contrary to the Gaussian wave packet in quantum molecular dynamics~\cite{AicPR202}, we do not attach any physical meaning to the form factor $S(\vec{r} - \vec{r}')$ and the smoothed LHV density $\tilde{\rho}$, and regard them as numerical techniques introduced in the test-particle approach so that we can obtain well-defined densities and mean fields.

In order to compensate for effects caused by the smearing of the local density due to the form factor, an additional gradient term should appear in the local density based on the following argument.
The LHV density $\tilde{\rho}$ can be regarded as the convolution of the realistic local density,
\begin{equation}
    \tilde{\rho}(\vec{r}) = \int\rho(\vec{r}')S(\vec{r} - \vec{r}')d\vec{r}'.
\end{equation}
To express $\rho$ in terms of $\tilde{\rho}$, formally we have
\begin{align}
    \rho(\vec{r}) & = \int\tilde{\rho}(\vec{r}')S^{-1}(\vec{r}' - \vec{r})d\vec{r}'\notag\\
    & = \int\Big[\sum_{n = 0}^{\infty}\frac{1}{n!}\nabla^n\tilde{\rho}(\vec{r})(\vec{r}' - \vec{r})^n\Big]S^{-1}(\vec{r}' - \vec{r})d\vec{r}'\notag\\
    & \approx\tilde{\rho}(\vec{r}) + c\nabla^2\tilde{\rho}(\vec{r})\label{E:rho},
\end{align}
where $S^{-1}(\vec{r} - \vec{r}')$ is the inverse of $S(\vec{r} - \vec{r}')$ which satisfies
\begin{equation}
    \int S(\vec{r} - \vec{r}'')S^{-1}(\vec{r}'' - \vec{r}')d\vec{r}'' = \delta(\vec{r} - \vec{r}').
\end{equation}
The parameter $c$, defined as
\begin{equation}
    c \equiv \int\frac{1}{2}(\vec{r}' - \vec{r})^2S^{-1}(\vec{r}' - \vec{r})d\vec{r}',
\end{equation}
is a constant only depending on the form of $S$.

In the LHV method, the direct correction on $\tilde{\rho}(\vec{r})$ is not feasible since numerically the density in Eq.~(\ref{E:rho}) is not always positively defined.
In practice, to compensate the smearing of density due to the form factor, we introduce an additional gradient term $\tilde{E}^{[2]}\nabla^2\tilde{\rho}(\vec{r})$ into the Hamiltonian.
To demonstrate this, one need only substitute Eq.~(\ref{E:rho}) into Eq.~(\ref{E:H}), and after several necessary approximations, an additional term proportional to $c\tilde{\rho}(\vec{r})\nabla^2\tilde{\rho}(\vec{r})$ will show up in the Hamiltonian.
Though $\tilde{\rho}$ is not a constant, in the present work, for simplicity the additional gradient term is recast effectively into $\tilde{E}^{[2]}\nabla^2\tilde{\rho}$, with $\tilde{E}^{[2]}$ being a constant.
This is equivalent to replace the gradient term coefficient $E^{[2]}$ by $E^{[2]}$ $+$ $\tilde{E}^{[2]}$.
Since in principle the rms radius in exact ground state does not change with time, $\tilde{E}^{[2]}$ here is adjusted roughly to obtain the ground state rms radius evolution with the smallest oscillation.
It should be mentioned that this correction on gradient term only improves the stability of ground state evolution~(rms radius and radial density profile) slightly, and does not cause much difference on the result of collective motions in Sec.~\ref{S:R&D}.
Needless to say, in ideal cases with $N_{\rm E}$ approaching infinity and lattice spacing approaching zero, LHV local density will approach the realistic local density, and $\tilde{E}^{[2]}$ will become zero.

\begin{table}[!htb]
\centering
\caption{The gradient parameters, both $E^{[2]}$ and $\tilde{E}^{[2]}$ for SP$6$s, SP$6$m, SP$6$h, and MSL$1$ used in the present work.
The obtained binding energy and proton rms radius of \isotope[208]{Pb} of TF initialization and LHV calculations are also shown.}
\begin{tabular}{cccccc}
\hline\hline
 ~ & \rm{SP$6$s} & \rm{SP$6$m} & \rm{SP$6$h} & \rm{MSL$1$} & Exp. \\
\hline
 $E^{[2]}$~($\rm MeV\cdot fm^{5}$) & -250.0 & -200.0 & -150.0 & -250.0 & - \\
 $\tilde{E}^{[2]}$~($\rm MeV\cdot fm^{5}$) & -15.0 & -10.0 & -10.0 & -20.0 & - \\
 \\
 BE~($\rm MeV$) & $1637.2$ & $1669.7$ & $1654.5$ & $1632.7$ & $1636.4$\\
 $\sqrt{\langle r^2\rangle_p}$~($\rm fm$) & $5.48$ & $5.44$ & $5.40$ & $5.51$ & $5.45$\\
 $\rm BE$~($\rm MeV$) in LHV & $1557.2$ & $1585.1$ & $1565.1$ & $1553.5$ & -\\
 $\sqrt{\langle r^2\rangle_p}$~($\rm fm$) in LHV & $5.52$ & $5.49$ & $5.44$ & $5.56$ & -\\
\hline\hline
\end{tabular}\label{T:E2}
\end{table}

In Table~\ref{T:E2}, for the interactions used in the present work, namely SP$6$s, SP$6$m, SP$6$h, and MSL$1$, we list their parameters $E^{[2]}$ and $\tilde{E}^{[2]}$, as well as, for \isotope[208]{Pb}, the binding energy and proton rms radius based on TF initialization, i.e., the solution of Eq.~(\ref{E:GS}), and in the LHV calculations.
We note from the table that after the smearing of $S$ in LHV method, the total energy decreases and the rms radius of proton increases slightly.
However, this difference affects the stability of the ground state very little, as we will see in Sec.~\ref{S:GS}.

\subsection{\label{S:GRLHV}Nuclear giant resonance within the Vlasov equation}

We consider a perturbative excitation of the Hamiltonian at the initial time, i.e.,
\begin{equation}
    \hat{H}_{ex}(t) = \lambda\hat{Q}\delta(t),
\end{equation}
where $\hat{Q}$ is the excitation operator we are interested in and $\lambda$ is supposed to be small.
Within the linear response theory~\cite{Fet1971}, the response of the excitation operator $\hat{Q}$ as a function of time is given by
\begin{align}
    \Delta\langle\hat{Q}\rangle(t) & = \langle f|\hat{Q}|f\rangle(t) - \langle0|\hat{Q}|0\rangle(t)\notag\\
    & = -\frac{2\lambda\theta(t)}{\hbar}\sum_f|\langle f|\hat{Q}|0\rangle|^2{\rm sin}\frac{(E_f-E_0)t}{\hbar},
    \label{E:dQt}
\end{align}
where $|0\rangle$ is the ground state for unperturbed system, $|f\rangle$ is the energy eigenstate of the excited system, $E_0$ and $E_f$ are the eigen-energy of the system before and after excitation, respectively.

We define the strength function $S(E)$ as usual through
\begin{equation}
    S(E) \equiv \sum_f|\langle f|\hat{Q}|0\rangle\delta(E - E_f - E_0).
\end{equation}
The $S(E)$ can be expressed as the Fourier integral of $\Delta\langle\hat{Q}\rangle(t)$ by taking advantage of Eq.~(\ref{E:dQt}), i.e.,
\begin{equation}\label{E:S-Q}
    S(E) = -\frac{1}{\pi\lambda}\int_0^{\infty}dt\Delta\langle\hat{Q}\rangle(t){\rm sin}\frac{Et}{\hbar}.
\end{equation}
By evaluating the time evolution of the response of $\hat{Q}$ within the LHV method, we can obtain the strength function.

We assume $\hat{Q}$ is a one-body operator, which means it can be expressed as the sum of operators $\hat{q}$ acting on each nucleon, i.e.,
\begin{equation}
    \hat{Q} = \sum_i^A\hat{q}.
\label{E:Qq}
\end{equation}
The expectation of $\hat{Q}$ then can be calculated as follows,
\begin{align}
    \langle\hat{Q}\rangle = & \langle f|\hat{Q}|f\rangle\notag\\
    = & \int\langle f|\vec{r}_1\cdots\vec{r}_N\rangle\langle\vec{r}_1\cdots\vec{r}_N|\hat{Q}|\vec{r}_1'\cdots\vec{r}_N'\rangle\notag\\
    &\times\langle\vec{r}_1'\cdots\vec{r}_N'|f\rangle d\vec{r}_1\cdots d\vec{r}_Nd\vec{r}_1'\cdots d\vec{r}_N'\label{E:Qt},
\end{align}
where we have added two identity operators.
Considering the definition of the one-body density matrix,
\begin{equation*}
    \rho(\vec{r}_1,\vec{r}_1') = A\int\langle\vec{r}_1\vec{r}_2\cdots\vec{r}_N|\Phi\rangle\langle\Phi|\vec{r}_1'\vec{r}_2\cdots\vec{r}_N\rangle d\vec{r}_2\cdots d\vec{r}_N,
\end{equation*}
and combining with Eq.~(\ref{E:Qq}), we rewrite Eq.~(\ref{E:Qt}) as
\begin{equation}
    \langle\hat{Q}\rangle = \int\rho(\vec{r}_1',\vec{r}_1)\langle\vec{r}_1|\hat{q}|\vec{r}_1'\rangle d\vec{r}_1d\vec{r}_1'\label{E:Q2}.
\end{equation}
For convenience in the following we change the variables of the integral, $\vec{r}_1$ $=$ $\vec{r} + \frac{\vec{s}}{2}$ and $\vec{r}_1'$ $=$ $\vec{r} - \frac{\vec{s}}{2}$.
Since $f(\vec{r},\vec{p})$ is the Wigner transform of density matrix, in coordinate space the density matrix can be expressed as
\begin{equation}
    \rho\Big(\vec{r} - \frac{\vec{s}}{2},\vec{r} + \frac{\vec{s}}{2}\Big) = \int f(\vec{r},\vec{p}){\rm exp}\Big(i\frac{\vec{p}}{\hbar}\vec{s}\Big)d\vec{p}\label{E:Q3}.
\end{equation}
We define the Wigner transform of $\hat{q}$ in coordinate space,
\begin{equation}
    q(\vec{r},\vec{p}) \equiv \int{\rm exp}\Big(-i\frac{\vec{p}}{\hbar}\cdot\vec{s}\Big)q\Big(\vec{r}+\frac{\vec{s}}{2},\vec{r}-\frac{\vec{s}}{2}\Big)d\vec{s}\label{E:Q4},
\end{equation}
where
\begin{equation}
    q\Big(\vec{r} + \frac{\vec{s}}{2},\vec{r} - \frac{\vec{s}}{2}\Big) = \Big\langle\vec{r} + \frac{\vec{s}}{2}|\hat{q}|\vec{r} - \frac{\vec{s}}{2}\Big\rangle
\end{equation}
is the matrix element of $\hat{q}$ in coordinate space.
Substituting Eq.~(\ref{E:Q3}) and the inverse transform of Eq.~(\ref{E:Q4}) into Eq.~(\ref{E:Q2}), we reduce the expectation of $\hat{Q}$ into the following form,
\begin{equation}
    \langle\hat{Q}\rangle = \int f(\vec{r},\vec{p})q(\vec{r},\vec{p})d\vec{r}d\vec{p}\label{E:Q5}.
\end{equation}

In the LHV method, the $f(\vec{r},\vec{p})$ is replaced by $\tilde{f}(\vec{r},\vec{p})$ expressed in Eq. (20).
It can be demonstrated that the effect of the external excitation $\lambda\hat{Q}\delta(t-t_0)$ is to change
the positions and momenta of the test nucleons as follows~\cite{UrbPRC85}:
\begin{equation}\label{E:q}
    \vec{r}_i \longrightarrow \vec{r}_i + \lambda\frac{\partial q(\vec{r}_i,\vec{p}_i)}{\partial\vec{p}_i},\quad\quad \vec{p}_i \longrightarrow \vec{p}_i - \lambda\frac{\partial q(\vec{r}_i,\vec{p}_i)}{\partial\vec{r}_i}.
\end{equation}
For the isoscalar monopole and isovector dipole excitation, the specific form of Eq.~(\ref{E:q}) will be given in Secs.~\ref{S:MS} and \ref{S:DV}, respectively.

\section{\label{S:R&D}Result and discussion}

In the following, we are going to examine the ability of the present LHV method in dealing with the (near-)equilibrium nuclear dynamics.
Specifically, we will study the ground state evolution, isoscalar monopole and isovector dipole mode of \isotope[208]{Pb}.
Under such (near-)equilibrium states, most of the nucleon-nucleon collisions are blocked according to the Pauli exclusive principle, thus in principle the BUU equation with the absence of the collision term, or the Vlasov equation, is still applicable.

\subsection{\label{S:GS}Ground state evolution of finite nuclei}

\begin{figure}[!htb]
\centering
\includegraphics[width=0.95\linewidth]{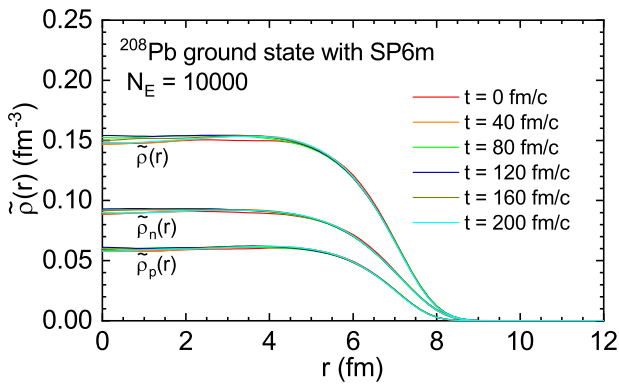}
\caption{The time evolution of radial density profile of ground state \isotope[208]{Pb} based on the present LHV method, with N$3$LO Skyrme pseudopotential SP$6$m up to $200~{\rm fm}/c$.}
\label{F:DP1}
\end{figure}

As mentioned in Sec.~\ref{S:INI}, the initial phase space information of a ground state nucleus is obtained self-consistently by varying the total energy with respect to nucleon radial density $\rho(r)$.
The initial coordinates of test nucleons are generated according to the solution of Eq.~(\ref{E:GS}), and the LHV density $\tilde{\rho}$ is obtain via Eq.~(\ref{E:rhoL}).
Shown in Fig.~\ref{F:DP1} is the time evolution of radial profile of LHV density for a single \isotope[208]{Pb} in ground state up to $200~{\rm fm}/c$, obtained from the present LHV method with $N_{\rm E}$ $=$ $10000$ and time step $\Delta t$ $=$ $0.4~{\rm fm}/c$ by using the N$3$LO Skyrme pseudopotential SP$6$m.
We notice from Fig.~\ref{F:DP1} that the LHV density approximates reasonably well the realistic ground state of the Vlasov equation, or the solution of Eq.~(\ref{E:GS}), since the profile of the LHV density only exhibits very small fluctuations.

\begin{figure}[!htb]
\centering
\includegraphics[width=0.95\linewidth]{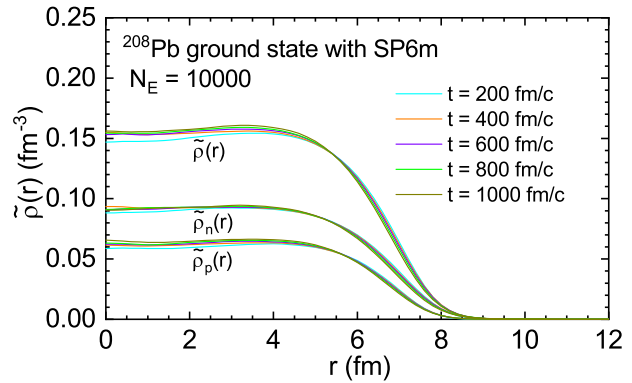}
\caption{Same as Fig.~\ref{F:DP1} but up to 1000~${\rm fm}/c$.}
\label{F:DP2}
\end{figure}

To see more clearly the stability of the ground state evolution of \isotope[208]{Pb} within the present LHV method, the ground state evolution is continued up to $1000~{\rm fm}/c$, and we present in Fig.~\ref{F:DP2} the radial profiles of LHV density with a time interval of $200~{\rm fm}/c$.
Again, only small fluctuations are observed in Fig.~\ref{F:DP2}, which indicates that the present LHV method is capable of studying long-time nuclear processes, e.g., heavy-ion fusion reactions and nuclear spallation reactions.

\begin{figure}[!htp]
\centering
\includegraphics[width=1.0\linewidth]{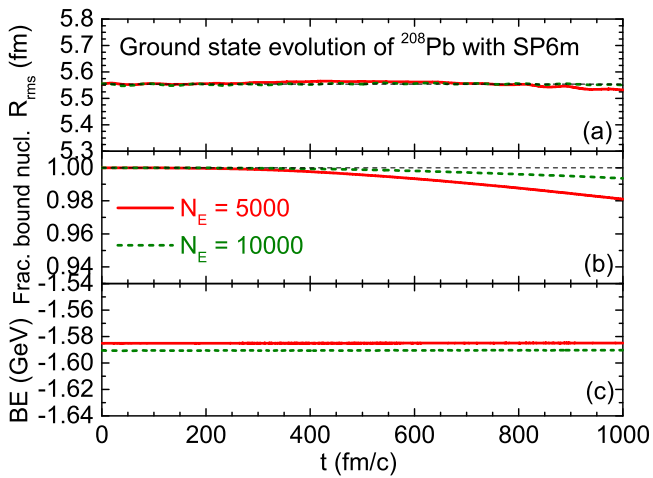}
\caption{Time evolution of (a)~rms radius, (b) the fraction of bound nucleons and (c) binding energy of ground state \isotope[208]{Pb} with N$3$LO Skyrme pseudopotential SP$6$m up to $1000~{\rm fm}/c$.
Calculations are performed with time step $\Delta t$ $=$ $0.4~{\rm fm}/c$, and $N_{\rm E}$ $=$ $5000$ and $10000$, respectively.}
\label{F:TE}
\end{figure}

Apart from the radial density profile, other properties concerning the stability of the LHV method are also examined.
To that end, the time evolution of rms radius, fraction of bound nucleons and binding energy are presented in Fig.~\ref{F:TE}.
The calculations are performed with time step $\Delta t$ $=$ $0.4~{\rm fm}/c$, and $N_{\rm E}$ $=$ $5000$ and $10000$, respectively.
Free test nucleons are those who do not interact with other test nucleons~(which means their form factors $S$ do not overlap), and they are excluded in calculating the fraction of bound nucleon and rms radius.
We notice from Fig.~\ref{F:TE}(a) that both cases with different $N_{\rm E}$ give fairly stable evolution of rms radius.
For the $N_{\rm E}$ $=$ $5000$ case, the rms radius starts to decrease after about $800~{\rm fm}/c$.
This decrease is due to the evaporation of nucleons from the bound nuclei, which is demonstrated in Fig.~\ref{F:TE}(b).
Such evaporation of nucleons is inevitable in transport model simulations due to the limited precision in the numerical realization, whereas it can be suppressed by increasing $N_{\rm E}$, as can be seen in Fig.~\ref{F:TE}(b), though the result with $E_{\rm E}$ $=$ $5000$ is already satisfactory.
Owing to the advantage of lattice Hamiltonian framework, the binding energy of the given nucleus conserves almost exactly in both cases, as shown in Fig.~\ref{F:TE}(c).

Figure \ref{F:TE} indicates that the present LHV method is able to give fairly stable time evolution.
Due to the high efficiency of GPU parallel computing, including more ensembles in the LHV calculation becomes possible.
However, as a balance between computational resources and numerical accuracy, unless otherwise specified, the following calculations are performed with time step $\Delta t$ $=$ $0.4~{\rm fm}/c$ and $N_{\rm E}$ $=$ $5000$.

\subsection{\label{S:MS}Isoscalar monopole mode of \isotope[208]{Pb}}

During the past several decades, a lot of studies on the isoscalar giant monopole resonance~(ISGMR) of finite nuclei have been performed, since it provides the information about both the symmetric and asymmetric part of the nuclear matter incompressibility~\cite{YouPRL82,ShlPRC47,TLPRL99,PatPLB718,PatPLB726,GupPLB760}, which are fundamental quantities characterizing the EOS of nuclear matter.
In the experimental aspect, the isoscalar monopole mode is measured through scattering off nucleus with isoscalar light particles, and recent experiments have been performed with inelastic $\alpha$-particle and deuteron scatterings~\cite{TLPRL99,MonPRL100,PatPLB718,PatPLB726,VanPRL113,GupPLB760}.

In the one-body transport model point of view, the isoscalar monopole mode is regarded as a compressional breathing of nuclear fluid.
Such mode can be generated in the LHV framework through the following procedures.
For the isoscalar monopole mode, we have
\begin{equation}
    \hat{Q}_{\rm ISM} = \frac{1}{A}\sum_i^{\rm A}\hat{r}_i^2\label{E:QMS},\quad\hat{q}_{\rm ISM} = \frac{\hat{r}^2}{A},
\end{equation}
and thus according to Eq.~(\ref{E:Q4}), we obtain
\begin{equation}
    q_{\rm ISM}(\vec{r},\vec{p}) = \frac{\vec{r}^2}{A}.
\end{equation}
Note that the square root of the expectation value of $\hat{Q}_{\rm ISM}$ is the rms radius of the given nucleus.
According to Eq.~(\ref{E:q}), to obtain the isoscalar monopole mode, the initial phase space information of test nucleons are changed with respect to that of the ground state by
\begin{equation}
    \vec{p}_i\longrightarrow\vec{p}_i - 2\lambda\frac{\vec{r}_i}{A}.
\end{equation}
The spatial coordinators of test nucleons remain unchanged since the $q_{\rm ISM}$ is independent of momentum.

\begin{figure}[!hbt]
\centering
\includegraphics[width=1.0\linewidth]{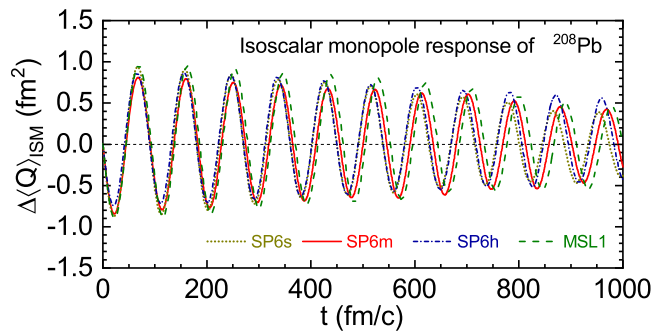}
\caption{The time evolution of $\Delta\langle\hat{Q}\rangle_{\rm ISM}$ of \isotope[208]{Pb} after a perturbation of $\hat{H}_{ex}(t)$ $=$ $\lambda\hat{Q}_{\rm ISM}\delta(t)$ with $\lambda$ $=$ $100~{\rm MeVfm^{-1}}/c$.
The results correspond to three N$3$LO Skyrme pseudo potentials, SP$6$s, SP$6$m, SP$6$h, and one conventional Skyrme interaction MSL$1$, respectively.}
\label{F:QMS}
\end{figure}

In Fig.~\ref{F:QMS} we show the time evolution of $\Delta\langle\hat{Q}\rangle_{\rm ISM}$ with three N$3$LO Skyrme pseudopotentials, namely, SP$6$s, SP$6$m, and SP$6$h, as well as one conventional Skyrme interaction MSL$1$.
The perturbation parameter $\lambda$ is chosen to be $100~{\rm MeVfm^{-1}}/c$ in the calculation.
We notice from the figure that the time evolution of $\Delta\langle \hat{Q}\rangle_{\rm ISM}$, or equivalently the rms radius, exhibits a very regular oscillation, and the rapid increase which is commonly observed in most BUU simulations using the conventional test particle method does not show up here.
Besides that, only a slight damping is observed in the oscillation, which is anticipated since the only damping mechanism in the Vlasov framework is Landau damping.
Landau damping is caused by one-body dissipation which is governed by a coupling of single-particle and collective motion.
It should be mentioned that in the RPA framework, the damping also only comes from one-body dissipation, since the coupling to more complex states like two-particle two-hole~($2p$-$2h$) states is missing in RPA~\cite{BerRMP55}.
In the semiclassical framework, effects analogous to the $2p$-$2h$ excitation can be included via a nucleon-nucleon collision term~\cite{BurNPA476}, and the width of the strength indeed increases due to the inclusion of the collision term~\cite{GaiPRC81}, but this is beyond the scope of the present work, and will be pursued in the future.

The strength function is obtained based on the time evolution of the response presented in Fig.~\ref{F:QMS} and Eq.~(\ref{E:Qq}).
When calculating the strength function, the response $\Delta\langle \hat{Q}\rangle(t)$ is multiplied by an exponential attenuation $e^{-\gamma t/2\hbar}$, which is a common practice in many giant resonance calculations within the BUU model~\cite{UrbPRC85,HKPRC95}.
The reason for introducing such an attenuation is to avoid oscillations in the Fourier transforms of Eq.~(\ref{E:S-Q}), which is caused by the finite period of the evolution.
In this work the attenuation parameter $\gamma$ is set to be $2~\rm MeV$, as well as in Sec.~\ref{S:DV} for the isovector dipole mode.
However, this exponential attenuation will not affect the peak energy of the strength, on which we mainly focus.

\begin{figure}[!htb]
\centering
\includegraphics[width=1.0\linewidth]{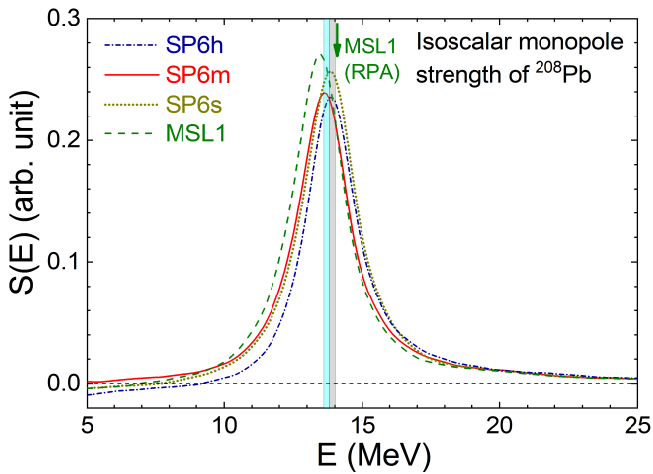}
\caption{Strength function of isoscalar monopole mode calculated based on the LHV method with SP$6$s, SP$6$m, SP$6$h, and MSL$1$.
The experimental peak energy and that from RPA calculation with MSL1 are also included for comparison.}
\label{F:SMS}
\end{figure}

The obtained strength functions of the isoscalar monopole mode with SP$6$s, SP$6$m, SP$6$h, and MSL$1$ are presented in Fig.~\ref{F:SMS}.
The grey band represents the peak energy of $13.91\pm0.11~\rm MeV$ from inelastic $\alpha$-scattering off \isotope[208]{Pb} performed at TAMU~\cite{YouPRL82} while the cyan band of $13.7\pm0.1~\rm MeV$ represents that from the Research Center for Nuclear Physics~(RCNP)~\cite{PatPLB726}.
We notice from the figure that all these interactions give peak energies consistent with that of the experiment, which is a consequence of the proper nuclear incompressibility of the given interactions, as shown in Table~\ref{T:CPs} by $K_0$.
In order to compare the results from the LHV method with that from different approaches, we calculate the strength function of MSL$1$ based on RPA.
The Skyrme-RPA code by Colo {\it et al.}~\cite{ColCPC184} is employed.
The obtained peak energy is indicated by a green arrow.
The peak energies of MSL$1$ with different approaches are generally comparable, and the small discrepancy comes from the difference between semi-classical and quantum nature.

\subsection{\label{S:DV}isovector dipole mode of \isotope[208]{Pb}}

The isovector giant dipole resonance~(IVGDR) of finite nuclei is the oldest known nuclear collective motion.
Systematic experimental studies on IVGDR based on photo-nuclear reactions can be traced back to more than forty years ago~\cite{BerRMP47}.
Recent measurements on isovector dipole response have been performed based on inelastic proton scattering at RCNP for \isotope[48]{Ca}~\cite{BirPRL118}, \isotope[120]{Sn}~\cite{HasPRC92}, and \isotope[208]{Pb}~\cite{TamPRL107}, as well as by using Coulomb excitation in inverse kinematics at GSI for \isotope[68]{Ni}~\cite{RosPRL111}.
It is interesting to mention that in recent years a low-lying mode called pygmy dipole resonance~(PDR) has attracted a lot of attention both experimentally~\cite{RyePRL89,AdrPRL95,WiePRL102,EndPRL105} and theoretically~\cite{CarPRC81,UrbPRC85,BarPRC88,BarEPJD68}.
It is well known from theoretical studies based on various models that the PDR, IVGDR, and electric dipole polarizability $\alpha_D$ which is dominated by these isovector dipole modes, provide sensitive probes of the density dependence of the nuclear symmetry energy~\cite{YilPRC72,TriPRC77,PiePRC85,RocPRC88,ZZPRC92,RocPRC92,HKPRC95}.

Within the LHV method, the external perturbation for the isovector dipole mode can be expressed in the following form:
\begin{equation}
    \hat{Q}_{\rm IVD} = \frac{N}{A}\sum_i^{\rm Z}\hat{z}_i - \frac{Z}{A}\sum_i^{\rm N}\hat{z}_i\label{E:QDV},
\end{equation}
which is defined so that the center of mass of the nucleus stays at rest.
Similar to the case in the isoscalar monopole mode, the excited system is obtained by changing the initial phase space coordinates of test nucleons according to
\begin{equation}
    p_z \longrightarrow
    \begin{cases}
         &p_z - \lambda\frac{N}{A},\quad \rm for~protons,\\
         &p_z + \lambda\frac{Z}{A},\quad \rm for~neutrons.
    \end{cases}
\end{equation}

\begin{figure}[!hbt]
\centering
\includegraphics[width=1.0\linewidth]{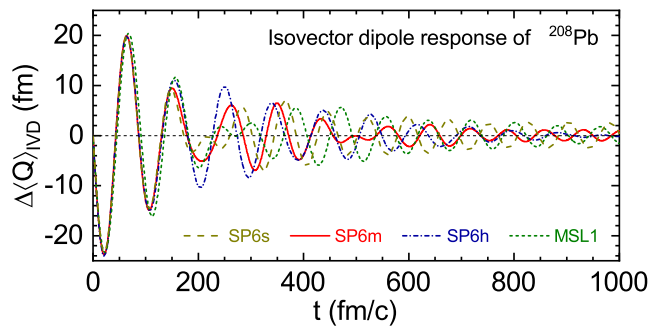}
\caption{Same as Fig.~\ref{F:QMS} but for isovector dipole mode.}
\label{F:QDV}
\end{figure}

\begin{figure}[!hbt]
\centering
\includegraphics[width=1.0\linewidth]{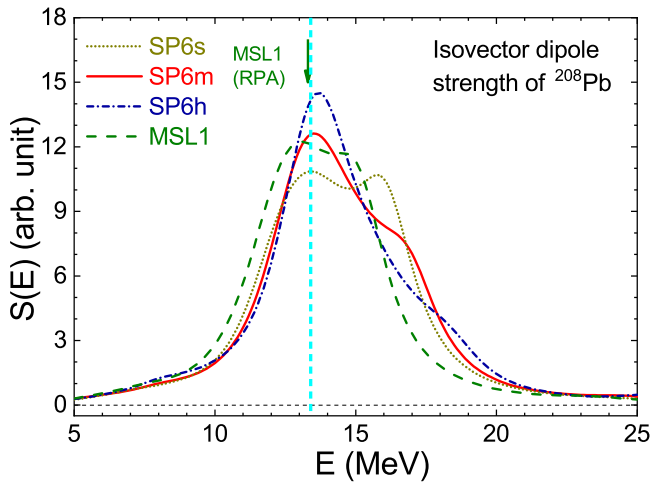}
\caption{Same as Fig.~\ref{F:SMS} but for isovector dipole mode.}
\label{F:SDV}
\end{figure}

After initializing and exciting the nucleus, we observe in Fig.~\ref{F:QDV} the damped oscillations of the isovector dipole response in \isotope[208]{Pb} with SP$6$s, SP$6$m, SP$6$h, and MSL$1$.
In the isovector dipole case, $\lambda$ is set to be $25~{\rm MeV}/c$.
Similar to the isoscalar monopole case, in Fig.~\ref{F:SDV} the corresponding strength functions of the isovector dipole response are displayed.
The vertical cyan line represents the experimental peak energy of $13.4~\rm MeV$ obtained from $\isotope[208]{Pb}(p,p')$ reaction performed at RCNP~\cite{TamPRL107} and the green arrow indicates the peak energy from the RPA calculation in MSL$1$.
Since these interactions give a reasonable description for the empirical isospin dependent behaviors in (sub)saturation density, as indicated in Table~\ref{T:CPs}, the peak energy from different interactions is consistent with experimental data.
The peak energies of MSL$1$ from the LHV method and the RPA calculation are comparable as well.

In Fig.~\ref{F:QDV}, the obtained responses with SP$6$s, SP$6$m, and MSL$1$ show clearly nonsingle frequency behavior.
This is also evident in Fig.~\ref{F:SDV} from the additional strength or even extra bumps in the high energy region.
Further tests show that the bump of the high energy part is related to the magnitude of the gradient terms in Eq.~(\ref{E:Hgrad}), which provides an additional restoring force out of phase with that from the local part.
This indicates that the effect of the gradient terms, usually omitted in one-body transport models, should not be overlooked.
We notice that such high-energy bumps or peaks have also been observed in previous studies based on one-body transport models with gradient terms~\cite{HZEPJWC117,HZPRC94}.
One possible reason is that the larger the gradient term, the easier for the isovector dipole excitation to invoke other modes, and other modes then react upon the isovector dipole mode and thus cause the additional high-energy strength.
In principle, such nonlinear modes can be avoided if the perturbation parameter $\lambda$ is small enough~\cite{SimPRC68}. However, this is impracticable in one-body transport model simulations due to the limited precision in the numerical realization.
Since the high-energy bump or peak is absent in experimental data~\cite{TamPRL107}, we attribute it to a numerical problem, and further investigation is needed in the future.

\section{\label{S:S&O}summary and outlook}
We have developed a one-body transport model by employing the lattice Hamiltonian method to solve the Vlasov equation with nuclear mean-field obtained from the N$3$LO Skyrme pseudopotential.
The ground states of nuclei are obtained with the same interactions through varying the total energy with respect to the nucleon density distribution.
Owing to the self-consistent treatment of initial nuclear ground state and the exact energy conservation of the LHV method, the present framework for solving the Vlasov equation exhibits very stable nuclear ground state evolution.
As a first application of the new LHV method, we have calculated the isoscalar monopole and isovector dipole modes of finite nuclei.
The obtained peak energies of ISGMR and IVGDR in \isotope[208]{Pb} with the N$3$LO Skyrme pseudopotentials are consistent with the experimental data.
In addition, the use of the Skyrme interaction enables us to compare the LHV results with that from conventional nuclear structure method, i.e., the RPA calculation, and the obtained peak energies of ISGMR and IVGDR in \isotope[208]{Pb} have been shown to be comparable in the LHV method and the RPA calculation.

Our results have demonstrated the capability of the present LHV method in dealing with the ground state of finite nuclei and the near-equilibrium nuclear dynamics.
The present work provides a solid foundation not only for long-time Vlasov calculation of low energy heavy-ion reactions, but also for the BUU calculation with a nucleon-nucleon collision term.
Based on the future lattice Hamiltonian BUU method, one can use the Skyrme pseudopotentials to simultaneously explore both the structure properties of finite nuclei and heavy-ion collisions at intermediate to high energies.
Crosschecks for nuclear effective interactions and thus for the EOS of asymmetric nuclear matter from the nuclear structure and heav-ion collisions thus become possible.
Such studies are in progress and will be reported elsewhere.

\section*{Acknowledgments}
We thank P. Danielewicz and B.-A. Li for helpful discussions,
and M. Gao and X. Zhang for the maintenance of the GPU severs.
This work was supported in part by the National Natural Science
Foundation of China under Grant No. 11625521, the Major State Basic Research
Development Program (973 Program) in China under Contract No.
2015CB856904, the Program for Professor of Special Appointment (Eastern
Scholar) at Shanghai Institutions of Higher Learning, Key Laboratory
for Particle Physics, Astrophysics and Cosmology, Ministry of
Education, China, and the Science and Technology Commission of
Shanghai Municipality (11DZ2260700).

\bibliography{GR-LHV}

\end{document}